\documentclass[11pt,a4paper]{article}
\usepackage{amsmath,amssymb,graphicx}
\usepackage{mathrsfs}
\usepackage{hyperref}
\usepackage{bm}
\usepackage{multirow}
\usepackage{leftidx}
\usepackage[text={17cm,24.5cm},centering]{geometry}
\usepackage{textcomp}
\usepackage{wasysym}
\numberwithin{equation}{section}

\newcommand{\be}{\begin{equation}}
\newcommand{\bea}{\begin{eqnarray}}
\newcommand{\eea}{\end{eqnarray}}
\newcommand{\ba}{\begin{array}}
\newcommand{\ea}{\end{array}}
\newcommand{\ee}{\end{equation}}
\newcommand{\ml}{\mathcal}
\newcommand{\ep}{\epsilon}
\newcommand{\de}{\delta}
\newcommand{\al}{\alpha}

\newcommand{\ga}{\gamma}
\newcommand{\ti}{\tilde}
\newcommand{\la}{\lambda}
\newcommand{\ihalf}{\frac{i}{2}}
\newcommand{\matr}[2]{\left(\begin{array}{#1}#2\end{array}\right)}
\newcommand{\nn}{\nonumber}

\def\({\left(}
\def\){\right)}
\def\[{\left[}
\def\]{\right]}
\begin{document}
\title{\textbf{Finite-size Effect for Dyonic Giant Magnons in $CP^3_{\beta}$ }}
\author{Hui-Huang Chen$^{a,b}$\footnote{chenhh@ihep.ac.cn}~,
Jun-Bao Wu$^{c,d,e}$\footnote{junbao.wu@tju.edu.cn}~
}
\date{}

\maketitle

\vspace{-10mm}

\begin{center}
{\it
$^{a}$Institute of High Energy Physics, and Theoretical Physics Center for Science
Facilities, Chinese Academy of Sciences, 19B Yuquan Road, Beijing 100049, P.~R.~China\\
$^{b}$ University of Chinese Academy of Sciences, 19A Yuquan Road, Beijing 100049, P.~R.~China\\
$^{c}$ School of Science, University of Tianjin, 92 Weijin Road, Tianjin 300072, P.~R.~China\\
$^{d}$School of Physics, Beihang University, 37 Xueyuan Road, Beijing 100191, P.~R.~China\\
$^{e}$Center for High Energy Physics, Peking University, 5 Yiheyuan Road, Beijing 100871, P.~R.~China}

\vspace{10mm}
\end{center}

\begin{abstract}
  We studied the finite-size giant magnons in $\text{AdS}_4\times\text{CP}^3_{\beta}$ background using the classical spectral curve constructed in this paper. We computed the finite-size corrections to the dispersion relations for the $RP^3$ giant magnons using our twisted algebraic curve based on the method proposed in \cite{Lukowski:2008eq}, in which the authors computed the finite-size corrections of giant magnons in $\text{AdS}_4\times\text{CP}^3$ by introducing a finite-size resolvent $G_{\text{finite}}(x)$. We obtained exactly the same result as in \cite{Ahn:2011nm}, where a totally different approach was used.
\end{abstract}
\thispagestyle{empty}
\newpage

\tableofcontents

\section{Introduction}
The gauge/gravity duality conjecture proposed in \cite{Maldacena:1997re} is a striking and thought-inspiring idea, which relates a large $N$ gauge theory to a certain string theory in such a non-trivial way that makes it very hard to prove. Since integrable structures were found in both sides in $\text{AdS}_5/\text{CFT}_4$ and $\text{AdS}_4/\text{CFT}_3$ dual pairs \cite{Minahan:2002ve},\cite{Bena:2003wd},\cite{Minahan:2008hf},\cite{Arutyunov:2008if}, integrability techniques can help us to understand and to test this conjecture qualitatively. See \cite{Beisert:2010jr} for a review. \\
\par Investigating such dualities without or with less supersymmetry is very important. One approach is to make marginal deformations in the field theory side. The $\beta$-deformation of $\text{AdS}_5/\text{CFT}_4$ is such an example, which in the field theory it corresponds to an exact marginal deformation \cite{Leigh:1995ep}, while the deformed background in the string theory can be generated by $\text{TsT}$ transformations \cite{Lunin:2005jy}. People have also investigated the three-parameter deformation which is also called $\ga$-deformation as a generalization of $\beta$-deformation \cite{Berenstein:2004ys},\cite{Beisert:2005if}. Another interesting example is the duality between type IIA string theory on $\text{AdS}_4\times\text{CP}^3_{\ga}$ and $\ga$-deformed ABJM theory \cite{Imeroni:2008cr}. This is a three-parameter deformation which breaks supersymmetry completely, while here we are interested in the $\beta$-deformed theory with $\ga_1=\ga_2=0,\ga_3=\beta$ which preserves $\ml{N}=2$ supersymmetry.\footnote{We always assume that the deformation parameters are real, since otherwise generically integrability will be broken \cite{Giataganas:2013dha}.} \\
\par In our recent paper \cite{Chen:2016geo}, the integrable structure of $\ga$-deformed ABJM theory was investigated in detail. The twist matrices were obtained in various bases. In this paper, we will try to calculate the finite-size effect
for giant magnons in $\text{AdS}_4\times\text{CP}^3_{\beta}$ using classical spectral curve. In order to compare our results with \cite{Ahn:2011nm}, we had better to use different charges from our previous choice in \cite{Chen:2016geo}. These charges are listed in Table~\ref{1}. These two choices can be easily related through non-singular linear transformations, so they are equivalent.  The twist matrix still has the form (in the $su_2$ grading)
\be
\textbf{A}=\frac12\matr{c|ccccc}
 {
 0 &\de_1-\de_3 & 0 & \de_1-\de_3 & -\de_1+2 \de_2+\de_3 & -\de_1-2 \de_2+\de_3 \\
\hline
 \de_3-\de_1 & 0 & 0 & \de_3-\de_1 & 2 \de_1 & -2 \de_3 \\
 0 & 0 & 0 & 0 & 0 & 0 \\
 \de_3-\de_1 & \de_1-\de_3 & 0 & 0 & \de_1+2\de_2+\de_3 & -\de_1-2 \de_2-\de_3 \\
 \de_1-2\de_2-\de_3 & -2\de_1 & 0 & -\de_1-2\de_2-\de_3 & 0 & 2\de_1+4\de_2+2\de_3 \\
 \de_1+2\de_2-\de_3 & 2\de_3 & 0 & \de_1+2\de_2+\de_3 & -2\de_1-4\de_2-2\de_3 & 0 \\
},
\ee
but with the relation between $\de_i$'s and $\ga_i$'s different,
\be
\ga_1=-\de_2,\quad\ga_2=-\de_1-\de_2-\de_3,\quad\ga_3=\de_3-\de_1.
\ee
\par We make natural assumption that the type IIA string theory on $\text{AdS}_4\times\text{CP}^3_{\ga}$
is integrable based on previous results in the field theory side. Its classical integrability can be proven similarly to the studies in \cite{Frolov:2005dj}.  Although we did not make  explicit calculations, we expect that the $\text{AdS}_4\times\text{CP}^3$ type IIA string in the $\ga$-deformed background with boundary condition $\varphi_i(2\pi)-\varphi_i(0)=2\pi m_i$\footnote{These $m_i$'s are not necessary to be integers when we consider giant magnons.} is equivalent to string with twisted boundary conditions $\ti{\varphi}_i(2\pi)-\ti{\varphi}_i(0)=2\pi(m_i-\ga_j\ep_{ijk}J_k)$ in the undeformed theory based on a general analysis on $\text{TsT}$ transformation, where $\varphi_i$'s are the three $U(1)$ directions $\varphi_1,\varphi_2,\psi$ listed in Table~\ref{1}, and $J_i$'s are the corresponding conserved angular momentum.\\
\par The dispersion relation for $RP^3$ giant magnons in $\text{AdS}_4\times\text{CP}^3_{\beta}$ has been calculated in \cite{Schimpf:2009rk}, and the finite-size corrections were obtained in \cite{Ahn:2011nm} by searching for the needed classical string solution directly. In this paper, we use the classical spectral curve method to calculate these quantities and to compare our results to the ones of \cite{Ahn:2011nm}. Finally, we find they match perfectly. \begin{table}\label{1}
  \centering
  \begin{tabular}{|c|c|c|c|c|c|c|c|c|}
  \hline
  Fields&$Y^1$&$Y^2$&$Y^3$&$Y^4$&$\Psi_1$&$\Psi_2$&$\Psi_3$&$\Psi_4$\\
  \hline
  $U(1)_{\varphi_1}$&$0$&$\frac{1}{2}$&$-\frac{1}{2}$&$0$ &$0$&$-\frac{1}{2}$&$\frac{1}{2}$&$0$ \\
  \hline
  $U(1)_{\varphi_2}$&$\frac{1}{2}$ & $0$ & $0$&$-\frac{1}{2}$&$-\frac{1}{2}$&$0$&$0$&$\frac{1}{2}$\\
  \hline
  $U(1)_{\psi}$&$-\frac{1}{2}$&$\frac{1}{2}$&$\frac{1}{2}$&$-\frac{1}{2}$&$\frac{1}{2}$&$-\frac{1}{2}$& $-\frac{1}{2}$&$\frac{1}{2}$ \\
  \hline
  \end{tabular}
  \caption{Charges of fields under three $U(1)$ generators of $SU(4)_R$.}
\end{table}
\section{$\ga$-deformed ABA at strong coupling}
In this section we briefly review the $\ga$-deformed $\text{AdS}_4/\text{CFT}_3$ asymptotic Bethe ansatz(ABA) equations and introduce some notations which will be useful in the following sections. In this section, we use the same notation as our previous paper \cite{Chen:2016geo}. Let's first define some useful functions as
\bea
&&R_l^{(\pm)}=\prod_{j=1}^{K_l}\(x(u)-x^{\mp}_{l,j}\)\,,R_l=\prod_{j=1}^{K_l}\(x(u)-x_{l,j}\)\,,\nn\\
&&B_l^{(\pm)}=\prod_{j=1}^{K_l}\(\frac{1}{x(u)}-x^{\mp}_{l,j}\)\,,B_l=
\prod_{j=1}^{K_l}\(\frac{1}{x(u)}-x_{l,j}\)\,,\nn\\
&&Q_l=\prod_{j=1}^{K_l}(u-u_{l,j})\,,S_l=\prod_{j=1}^{K_l}\sigma_{BES}\(x(u),x_{l,j}\)\,,
\eea
and the functions with no subscript mean a product of type-$4$ and type-$\bar4$ ones: $R=R_4R_{\bar4},B=B_4B_{\bar4}$. We also introduce the excitation number vector $\textbf{K}=(L|K_1,K_2,K_3,K_4,K_{\bar4})$ as before,
then the $\ga$-deformed ABA can be written as (in the $su_2$ grading)
\bea\label{TwistedABA}
&&e^{-2\pi i(\textbf{AK})_1}=e^{-i\ml{Q}_1}\frac{Q_2^+B^{(-)}}{Q_2^-B^{(+)}}\bigg|_{u=u_{1,k}}\,,\nn\\
&&e^{-2\pi i(\textbf{AK})_2}=-\frac{Q_1^+Q_2^{--}Q_3^+}{Q_1^-Q_2^{++}Q_3^-}\bigg|_{u=u_{2,k}}\,,\nn\\
&&e^{-2\pi i(\textbf{AK})_3}=\frac{Q_2^+R^{(-)}}{Q_2^-R^{(+)}}\bigg|_{u=u_{3,k}}\,,\nn\\
&&e^{-2\pi i(\textbf{AK})_4}=-e^{-ip(u_{4,k})L}
\frac{Q_4^{++}B_1^-R_3^-}{Q_4^{--}B_1^+R_3^+}S_4S_{\bar4}\bigg|_{u_{4,k}}\;,\nn\\
&&e^{-2\pi i(\textbf{AK})_{\bar4}}=-e^{-ip(u_{\bar4,k})L}
\frac{Q_{\bar4}^{++}B_1^-R_3^-}{Q_{\bar4}^{--}B_1^+R_3^+}S_4S_{\bar4}\bigg|_{u_{\bar4,k}}\;,
\eea
where the $x^\pm$ are Zhukowsky variables,
\be\label{X}
x+\frac{1}{x}=\frac{u}{h(\lambda)}\;\;,\;\;x^\pm+\frac{1}{x^\pm}=\frac{1}{h(\lambda)}\(u\pm\ihalf\) \,.
\ee
and we have already used the general notation
\be
f^{\pm}\equiv f(u\pm i/2),\, f^{\pm\pm}\equiv f(u\pm i), \, f^{[\pm a]}\equiv f(u\pm ia/2).
\ee
The spectrum of all conserved charges is given by the momentum carrying roots $u_4$ and $u_{\bar4}$ alone from,
\be
\ml{Q}_n=\sum_{j=1}^{K_4}\bold{q}_n(u_{4,j})+\sum_{j=1}^{K_{\bar4}}\textbf{q}_n(u_{\bar4,j}),\quad
\bold{q}_n=\frac{i}{n-1}\(\frac{1}{(x^+)^{n-1}}-\frac{1}{(x^-)^{n-1}}\)
\ee
The conserved momentum reads
\be
p=\ti{p}+2\pi(\textbf{AK})_0\,,
\ee
where
\be
\ti{p}=\ml{Q}_1=-i\sum_{j=1}^{K_4}\log\frac{x_{4,j}^+}{x_{4,j}^-}
-i\sum_{j=1}^{K_{\bar4}}\log\frac{x_{\bar4,j}^+}{x_{\bar4,j}^-}.
\ee
At weak coupling, $h({\lambda})\simeq \lambda$, and at strong coupling, $h({\lambda})\simeq \sqrt{\lambda/2}$.
The BES dressing phase $\sigma_{BES}$ behaves like:
\be
\sigma_{BES}(u,v)\rightarrow 1,\quad \text{when}\;\lambda\rightarrow 0,
\ee
\be
\sigma_{BES}(u_k,u_j)\rightarrow \frac{1-1/x^-_kx^+_j}{1-1/x^+_kx^-_j}\(\frac{1-x^-_kx^-_j}{1-x^+_kx^-_j}\frac{1-x^+_kx^+_j}{1-x^-_kx^+_j}\)^{i(u_j-u_k)},\;\text{when}\;\lambda\rightarrow\infty\,.
\ee
In the scaling limit $u_{a,j}\simeq L\simeq\sqrt{\la}\simeq K_a\gg1$, we find
\be
x^{\pm}=x(u\pm\ihalf)=x\pm\ihalf\al(x)+\ml{O}(\frac{1}{\lambda}),
\ee
where
\be
\al(x)=\frac{1}{2g}\frac{x^2}{x^2-1},
\ee
and $g$ is defined through $g\equiv\sqrt{\lambda/8}$.
It is convenient to introduce the resolvents:
\be
G_a(x)=\sum_{j=1}^{K_a}\frac{\al(x_{a,j})}{x-x_{a,j}},\quad H_a(x)=\sum_{j=1}^{K_a}\frac{\al(x)}{x-x_{a,j}},\quad \bar{G}_a(x)=G_a(1/x),\quad \bar{H}_a(x)=H_a(1/x).
\ee
We can recover the conserved charges $\ml{Q}_n$ from these resolvents:
\be
G_4(x)+G_{\bar4}(x)=-\sum_{n=0}^{\infty}\ml{Q}_{n+1}x^n.
\ee
\section{The classical spectral curve of $\ga$-deformed $\text{AdS}_4/\text{CFT}_3$}
In this section, we try to obtain the classical spectral curve from the generating functional \cite{Gromov:2009at},\cite{Kim:2014vka}. In the context of $\ga$-deformation, the twisted generating functional is needed. The twisted generating functional for $\ga$-deformed $\text{AdS}_4/\text{CFT}_3$ has been constructed in \cite{Chen:2016geo} and reads (we have used a different gauge from there)
\be
\ml{W}=\frac{1}{1-\frac{1}{\tau_1}\frac{B^{(-)-}Q_1^+}{B^{(+)-}Q_1^-}D}
\(1-\frac{1}{\tau_2}\frac{Q_1^+Q_2^{--}}{Q_1^-Q_2}D\)
\(1-\tau_2\frac{Q_2^{++}Q_3^-}{Q_2Q_3^+}D\)\frac{1}{1-\tau_1\frac{Q_3^-R^{(+)+}}{Q_3^+R^{(-)+}}D}\,,
\ee
where
\be
\tau_1=e^{-\pi i((\de_3-\de_1)L+(\de_1-\de_3)K_1+(\de_1+2\de_2+\de_3)(K_4-K_{\bar4}))}
\,,\quad\tau_2=1.
\ee
In the scaling limit, we have the following expansions
\bea
&&\frac{B^{(-)-}Q_1^+}{B^{(+)-}Q_1^-}\simeq\exp[-i(-\frac{x^2\ml{Q}_1+x\ml{Q}_2}{x^2-1}-H_1-\bar{H}_1+\bar{H}_4+\bar{H}_{\bar4})],\nn\\
&&\frac{Q_1^+Q_2^{--}}{Q_1^-Q_2}\simeq\exp[-i(-H_1-\bar{H}_1+H_2+\bar{H}_2)],\nn\\
&&\frac{Q_2^{++}Q_3^-}{Q_2Q_3^+}\simeq\exp[-i(-H_2-\bar{H}_2+H_3+\bar{H}_3)],\nn\\
&&\frac{Q_3^-R^{(+)+}}{Q_3^+R^{(-)+}}\simeq\exp[-i(-\frac{\ml{Q}_1+x\ml{Q}_2}{x^2-1}+H_3+\bar{H}_3-H_4-H_{\bar4})].
\eea
At strong coupling, the generating functional becomes
\be
\ml{W}=\frac{(1-\la_1d)(1-\la_2d)}{(1-\la_3d)(1-\la_4d)},
\ee
where $\la_a=e^{-iq_a}$, with $q_a$ being the so called quasi-momenta.
We have refined the formal expansion parameter in the following way
\be
d=\exp[i(\frac{Lx/2g+x\ml{Q}_2}{x^2-1}+\bar{H}_1-\bar{H}_3)]D.
\ee
The $\text{AdS}_4\times\text{CP}^3_{\ga}$ classical string algebraic curve is a ten sheets Riemann surface parameterized by $\{e^{iq_a(x)},e^{-iq_a(x)}\}$. After some computations, we find
\bea
q_3(x)&=&\frac{Lx/2g-\ml{Q}_1}{x^2-1}-H_1-\bar{H}_3+\bar{H}_4+\bar{H}_{\bar4}+\phi_3,\nn\\
q_2(x)&=&\frac{Lx/2g+\ml{Q}_2x}{x^2-1}+H_2-\bar{H}_3-H_1+\bar{H}_2+\phi_2,\nn\\
q_1(x)&=&\frac{Lx/2g+\ml{Q}_2x}{x^2-1}+H_3-H_2-\bar{H}_2+\bar{H}_1+\phi_1,\nn\\
q_4(x)&=&\frac{Lx/2g-\ml{Q}_1}{x^2-1}+\bar{H}_1+H_3-H_4-H_{\bar4}+\phi_4,\nn\\
q_5(x)&=&H_4-H_{\bar4}+\bar{H}_4-\bar{H}_{\bar4}+\phi_5,
\eea
where $\phi_3=-i\log\tau_1+2\pi(\textbf{AK})_0$,$\phi_4=i\log\tau_1$,$\phi_1=\phi_2=0$,
and $q_{11-i}=-q_i,i=1,2,\cdots,5$. As discussed in \cite{Chen:2016geo}, we must also introduce the twist phases
\be
\tau_0=e^{\pi i(-2\de_2L-(\de_1+\de_3)K_1-(\de_1+2\de_2+\de_3)(K_3-K_4-K_{\bar4}))}\,,
\tau_{\bar0}=e^{\pi i(2\de_2L+(\de_1+\de_3)K_1+(\de_1+2\de_2+\de_3)(K_3-K_4-K_{\bar4}))}\,
\ee
such that $\phi_5=-i\log\tau_0$.
The relations between the conserved angular momentum and the excitation numbers read
\bea
&&J_1\equiv J_{\varphi_1}=\frac{q}{2}=\frac{1}{2}(K_4+K_{\bar4}-K_1-K_3),\nn\\
&&J_2\equiv J_{\varphi_2}=\frac{p_1+q+p_2}{2}=L-\frac{K_4+K_{\bar4}}{2}+\frac{K_3-K_1}{2},\nn\\
&&J_3\equiv J_{\psi}=\frac{p_2-p_1}{2}=K_4-K_{\bar4},
\eea
where we denote the $SU(4)$ Dynkin labels as $[p_1,q,p_2]$. See Appendix~\ref{appenb} for more information.
Using the above relations, we can write the twists $\textbf{AK}$ appearing in the ABA equations as
\be\label{TwistCharge}
\textbf{AK}=\(\begin{array}{c}
\ga_3J_1-\ga_1J_3\\
\frac{\ga_1-\ga_2}{2}J_3+\frac{\ga_3}{2}(J_2-J_1)\\
0\\
\frac{\ga_3}{2}(J_1+J_2)-\frac{\ga_1+\ga_2}{2}J_3\\
\frac{\ga_1}{2}(2J_2+J_3)+\frac{\ga_2}{2}(J_3-2J_1)-\frac{\ga_3}{2}(J_1+J_2)\\
-\frac{\ga_1}{2}(2J_2-J_3)+\frac{\ga_2}{2}(J_3+2J_1)-\frac{\ga_3}{2}(J_1+J_2)
\end{array}\).
\ee
With the help of the formulas given in Appendix~\ref{appena}, it's very easy to show that the twisted ABA eqs.~(\ref{TwistedABA}) are equivalent to the following equations in the scaling limit.
\bea
&&q_3(x+i0)-q_2(x-i0)=-\frac{\ml{Q}_1+x\ml{Q}_2}{x^2-1}-H_2-\bar{H}_2+\bar{H}_4+\bar{H}_{\bar4}
+\phi_3-\phi_2=2\pi n_{1}\quad x\in\ml{C}_1,\nn\\
&&q_2(x+i0)-q_1(x-i0)=-H_1+2\,/\kern-0.7em H_2-H_3-\bar{H_1}+2\bar{H}_2-\bar{H}_3
+\phi_2-\phi_1=2\pi n_{2}\quad x\in\ml{C}_2,\nn\\
&&q_1(x+i0)-q_4(x-i0)=\frac{\ml{Q}_1+x\ml{Q}_2}{x^2-1}-H_2+H_4+H_{\bar4}-\bar{H}_2
+\phi_1-\phi_4=2\pi n_3\quad x\in\ml{C}_3,\nn\\
&&q_4(x+i0)-q_5(x-i0)=\frac{Lx/2g-\ml{Q}_1}{x^2-1}+H_3-2\,/\kern-0.7em H_4+\bar{H}_1-\bar{H}_4+\bar{H}_{\bar4}+\phi_4-\phi_5=2\pi n_4\quad x\in\ml{C}_4,\nn\\
&&q_4(x+i0)+q_5(x-i0)=\frac{Lx/2g-\ml{Q}_1}{x^2-1}+H_3-2\,/\kern-0.7em H_{\bar4}+\bar{H}_1-\bar{H}_{\bar4}+\bar{H}_4+\phi_4+\phi_5=2\pi n_5\quad x\in\ml{C}_{\bar4}.\nn\\
\eea
\section{Spectral curve for giant magnons in $\text{CP}^3_{\beta}$}
In this section and the following, we will work with the $\beta$-deformation of $\text{AdS}_4/\text{CFT}_3$, where we take $\ga_1=\ga_2=0,\ga_3\equiv\beta$, and we label the Dynkin nodes as $(r,u,v)$ instead of $(3,4,\bar4)$.
We use the ansatz for solutions mostly in $\text{CP}^3_{\beta}$ \cite{Lukowski:2008eq}\footnote{Comparing with \cite{Lukowski:2008eq}, We flip the sign in front of all resolvents in the ansatz for the quasi-momenta $q_1, \cdots, q_4$. Only with this choice, the asymptotic behavior of the quasi-momenta (\ref{AsymBehav}) and the total momentum from the inversion symmetry (\ref{momentum}) are correctly produced. Accordingly, we also flip the sign in (\ref{InfiniteResolvent}) and (\ref{finiteresolvent}). The algebraic curve for giant magnons was also studied in \cite{Shenderovich:2008bs}.},
\bea\label{Ansatz}
&&q_1(x)=\frac{\al x}{x^2-1}+\phi_1+\Delta\phi_1,\nn\\
&&q_2(x)=\frac{\al x}{x^2-1}+\phi_2+\Delta\phi_2,\nn\\
&&q_3(x)=\frac{\al x}{x^2-1}-G_u(0)+\bar{G}_u(x)-G_v(0)+\bar{G}_v(x)-G_r(x)+G_r(0)-\bar{G}_r(x)+\phi_3+\Delta\phi_3,\nn\\
&&q_4(x)=\frac{\al x}{x^2-1}-G_u(x)-G_v(x)+G_r(x)-G_r(0)+\bar{G}_r(x)+\phi_4+\Delta\phi_4,\nn\\
&&q_5(x)=G_u(x)-G_u(0)+\bar{G}_u(x)-G_v(x)+G_v(0)-\bar{G}_v(x)+\phi_5+\Delta\phi_5,
\eea
where the twists $\phi_i$ and $\Delta\phi_i$ adding here are to be determined. Let's make some comments on these two kind of twists. The giant magnons are open string solutions with endpoints located on different places. If we want to treat them as closed strings, we have to consider $\mathbb{Z}_M$ orbifold of deformed  ABJM theory. The $\Delta\phi_i$'s incorporate this effect. While the existence of $\phi_i$ are due to that we are considering the $\beta$-deformed theory. The Dynkin labels of $SU(4)$ are related to the excitation numbers by
\be
\[\begin{array}{c}
p_1\\q\\p_2
\end{array}\]
=\[\begin{array}{c}
L-2K_u+K_r\\
K_u+K_v-2K_r\\
L-2K_v+K_r
\end{array}\].
\ee
While in this sector,
\be
\[\begin{array}{c}
J_1\\J_2\\J_3
\end{array}\]
=\[\begin{array}{c}
\frac{q}{2}\\\frac{p_1+q+p_2}{2}\\\frac{p_2-p_1}{2}
\end{array}\]
=\[\begin{array}{c}
\frac12(K_u+K_v-2K_r)\\
\frac12(2L-K_u-K_v)\\
K_u-K_v
\end{array}\].
\ee
In the last section, we have calculate the twists $\phi_1,\dots,\phi_5$, with the help of eq.~(\ref{TwistCharge}), we can write them in terms of the angular momentum. For the case at hand, we have
\be
\phi_1=\phi_2=0,
\phi_3=\pi\beta(J_1-J_2),
\phi_4=\pi\beta(J_1+J_2),
\phi_5=0.
\ee\\
\subsection{Fix the twists from orbifolding}
\par In this section we will mainly discuss the $RP^3$ giant magnon, which is the dyonic generalization of $RP^2$ giant magnon. We closely follow the treatment of \cite{Gromov:2008ie} (early work on treating giant magnon as closed string on orbifold includes
\cite{Astolfi:2007uz}-\cite{Grignani:2008te}). Let us temporarily put aside the deformation.  As mentioned above, giant magnons are open string solutions with non-periodic boundary conditions. For $RP^3$ magnon, we have
\bea
&&Y^1(\sigma=2\pi)=Y^1(\sigma=0)\exp(ip/2),\nn\\
&&Y^2(\sigma=2\pi)=Y^2(\sigma=0),\nn\\
&&Y^3(\sigma=2\pi)=Y^3(\sigma=0),\nn\\
&&Y^4(\sigma=2\pi)=Y^4(\sigma=0)\exp(-ip/2).
\eea
Formally identifying this open string as closed string leads us to consider $\mathbb{Z}_M$ orbifolding of ABJM theory. This theory appears when we
consider low energy effective theory of $N$ $M2$ branes put at $\mathbb{C}^4/(\mathbb{Z}_M\times \mathbb{Z}_{Mk})$ orbifold singularity \cite{c1}- \cite{Berenstein} with $\mathbb{Z}_M$ acting on $\mathbb{C}^4$
as
 \be (Y^1, Y^2, Y^3, Y^4)\to (\exp(2\pi m/M)Y^1, Y^2, Y^3, \exp(-2\pi m/M)Y^4).\ee
 Notice that this $\mathbb{Z}_M$ is inside $SU(4)$ R-symmetry group of ABJM theory.
It is easy to see that we should identify  $p/2$ with $2\pi m/M$.
 Then we have to set the charges of $Y^I$'s as \be s_I=(m, 0, 0, -m). \ee
From the analysis in \cite{Bai:2016pxs}, we have
\be
s_I=(t_2,t_1-t_2,-t_1+t_3,-t_3),
\ee
where $t_I$'s are parameters of the orbifold.
Comparing these two results, we find $t_1=t_2=t_3=m$. \\
Then the charges appearing in the $su_2$-grading orbifolding ABA equations are
\bea
\textbf{q}^+&=&(-t_2-t_3|0,-t_1,2t_1-t_2-t_3,-t_1+2t_2,-t_1+2t_3)\nn\\
&=&(-2m|m, 0, -m, m, m).
\eea
The phases from the orbifold are
\be
\exp(2\pi i \textbf{q}^+/M)=(e^{-ip}| e^{ip/2}, 1, e^{-ip/2}, e^{ip/2}, e^{ip/2}).
\ee
Notice that in the ABA equations, the phase due to deformation $\exp(-2\pi i (\textbf{A}\textbf{K})_j)$ should be multiplied by the phase due to new orbifolding
$\exp(-2 \pi i q_j^+/M)$, so the change of $\phi$'s due to this orbifolding should satisfy
\bea \Delta \phi_3-\Delta \phi_2&=&-p/2,\nn\\
\Delta \phi_2-\Delta \phi_1&=&0,\nn \\
\Delta \phi_1-\Delta \phi_4&=&p/2,\nn \\
\Delta \phi_4-\Delta \phi_5&=&-p/2,\nn \\
\Delta \phi_4+\Delta \phi_5&=&-p/2.
\eea
This leads to
\be \Delta\phi_1=\Delta\phi_2=\Delta\phi_5=0, \Delta\phi_3=\Delta\phi_4=-p/2,
\ee
exactly the same as the results given in \cite{Abbott:2010yb} (see also \cite{Abbott:2013mpa}).
Thus the asymptotic behaviour of the quasi-momentum when $x\rightarrow\infty$
\be\label{AsymBehav}
\(\begin{array}{c}
q_1\\q_2\\q_3\\q_4\\q_5
\end{array}\)\simeq \(\begin{array}{c}
0\\0\\\pi\beta(J_1-J_2)-p/2\\\pi\beta(J_1+J_2)-p/2\\0
\end{array}\)+\frac{1}{2gx}\(\begin{array}{c}
\Delta+S\\\Delta-S\\J_1+J_2\\J_2-J_1\\J_3
\end{array}\).
\ee
Comparing our ansatz eq.~(\ref{Ansatz}) with the asymptotic behaviors eq.~(\ref{AsymBehav}), we conclude $\al=\Delta/2g$ and $S=0$.\\
The total momentum condition from the Bethe ansatz now reads
\be \tilde{p}=2\pi n-2\pi (\textbf{A}\textbf{K})_0+p. \ee
Here $n\in \mathbb{Z}$, $p$ is the momentum of giant magnon and $\tilde{p}$ is defined as
\be
\tilde{p}=-i\sum_{j=1}^{K_4}\log\frac{x^+_{4,j}}{x^-_{4,j}}-i\sum_{j=1}^{K_{\bar4}}\log
\frac{x^+_{\bar4,j}}{x^-_{\bar4,j}}.
\ee
For $\beta$-deformed case, we have
\be
(\textbf{A}\textbf{K})_0=2\pi \beta J_1.
 \ee
And the above result is consistent with the results from inversion symmetry
\be \label{momentum} 2\pi n=q_3(1/x)+q_4(x)=\tilde{p}+\phi_3+\Delta \phi_3+\phi_4+\Delta \phi_4=\tilde{p}+2\pi \beta J_1-p. \ee
For our convenience, we can set $n=0$. In fact, any even $n$ gives the correct result. This require comes from the ambiguity of the definition of $\ti{p}$. Let's make this point more clear. $\ti{p}$ is defined up to some integer multiply $2\pi$. For `small giant magnon', any $n$ gives the same result. But for $RP^3$ magnon, this integer must be a even number, because we have two copies of `small giant magnon'.

\subsection{Infinite size dispersion relation}
\par Studying giant magnons using the classical spectral curve method first appears in \cite{Minahan:2006bd}, where the authors show that giant magnons correspond to logarithmic cuts in the algebraic curve language.
The giant magnons in $\text{CP}^3_{\beta}$ have various forms. But essentially they belong to two different classes: the `small' and `big' giant magnons and their dyonic generalizations. They can all be constructed by setting some of the resolvents in ansatz eq.~(\ref{Ansatz}) to be
\be\label{InfiniteResolvent}
G_{\text{mag}}(x)=i\log\(\frac{x-X^+}{x-X^-}\),
\ee
where $(X^+)^*=X^-$.
\par As a warm-up exercise, let's first consider the `small giant magnon' with
\be
G_u(x)=G_{\text{mag}}(x),\qquad G_v(x)=G_r(x)=0.
\ee
The charges can be read from the asymptotic behavior of this curve
\bea
\ti{p}=-i\log\frac{X^+}{X^-},\qquad J_1=-ig(X^+-X^-+\frac{1}{X^+}-\frac{1}{X^-}),\nn\\
J_2=\Delta+ig(X^+-X^--\frac{1}{X^+}+\frac{1}{X^-}),\qquad J_3=2J_1.
\eea
Then the dispersion relation can be find as
\be
\ml{E}=\Delta-J_2=\sqrt{J_1^2+16g^2\sin^2\frac{\ti{p}}{2}}\,,
\ee
where $\ti{p}=p-2\pi\beta J_1$.\\
\par Now we consider another kind of dyonic giant magnons in $\text{CP}^3_{\beta}$, which are also called pair of small giant magnons or $RP^3$ magnons. They can be constructed by putting a `small giant magnon' in each sector, $G_u(x)=G_v(x)=G_{\text{mag}}(x)$ and with $G_r(x)=0$.
From this setting, we obtain the charges as
\bea
\ti{p}=-2i\log\frac{X^+}{X^-},\qquad J_1=-2ig(X^+-X^-+\frac{1}{X^+}-\frac{1}{X^-}),\nn\\
J_2=\Delta+2ig(X^+-X^--\frac{1}{X^+}+\frac{1}{X^-}),\qquad J_3=0.
\eea
Thus we find the dispersion relation
\be
\ml{E}=\Delta-J_2=\sqrt{J_1^2+64g^2\sin^2\frac{\ti{p}}{4}}\,,
\ee
where $\ti{p}=p-2\pi\beta J_1$.
\section{Finite-size corrections}
The finite-size effects for giant magnons were usually computed through L\"uscher formula, which basically contain two different terms coming from two types of spacetime interpretation: the $\mu$-term and the F-term, which correspond to the leading classical corrections and the first quantum corrections respectively. The F-term is easily compute from classical spectral curve. In \cite{Lukowski:2008eq}, the authors proposed a method to compute the $\mu$-term from the algebraic curve, see also \cite{Sax:2008in}. As we don't not attend to construct the Drinfeld-Reshetikhin twist \cite{Ahn:2010ws} of the $\text{AdS}_4/\text{CFT}_3$ S-matrix in this short note, we will try to compute the finite-size corrections using the twisted classical spectral curve proposed above. The basic idea is that at finite size we can think the giant magnon solutions obtain a small square root cut tail in each ends of the logarithmic cut. Considering this, we use the finite-size resolvent \cite{Lukowski:2008eq}
\be\label{finiteresolvent}
G_{\text{finite}}(x)=2i\log\(\frac{\sqrt{x-X^+}+\sqrt{x-Y^+}}{\sqrt{x-X^-}+\sqrt{x-Y^-}}\),
\ee
where $Y^{\pm}$ are points shift by some small amount away from $X^{\pm}$:
\be
Y^{\pm}=X^{\pm}(1\pm i\de e^{\pm i\phi}).
\ee
As a simple check, when taking $\de=0$, this finite-size resolvent goes back to the infinite-size resolvent (\ref{InfiniteResolvent}).
Now we will study the finite-size $RP^3$ dyonic giant magnons using the method mentioned above. For this purpose, we set $G_u(x)=G_v(x)=G_{\text{finite}}(x)$ and with $G_r(x)=0$. We write
\be
X^{\pm}=re^{\pm i\ti{p}_0/4}.
\ee
The momentum of the giant magnon is
\be
p=\ti{p}+2\pi\beta J_1,
\ee
where
\be
\ti{p}=-4i\log\frac{\sqrt{X^+}+\sqrt{Y^+}}{\sqrt{X^-}+\sqrt{Y^-}}.
\ee
The expansion of $\ti{p}$ in small $\de$ is
\be
\ti{p}=\ti{p}_0+2\cos(\phi)\de+\frac{3}{4}\sin(2\phi)\de^2+\ml{O}(\de^3).
\ee
The non-trivial large $x$ asymptotic of this curve is
\bea
&&q_3(x)\simeq \pi\beta(J_1-J_2)-\frac{p}{2}+\frac{1}{2gx}\(\Delta+2ig\(\frac{2}{\sqrt{X^-Y^-}}-\frac{2}{\sqrt{X^+Y^+}}\)\),\nn\\
&&q_4(x)\simeq\pi\beta (J_1+J_2)-\frac{p}{2}+\frac{1}{2gx}(\Delta+2ig(X^+-X^-+Y^+-Y^-)),
\eea
from which we solve
\bea
J_1=ig\(\frac{2}{\sqrt{X^-Y^-}}-\frac{2}{\sqrt{X^+Y^+}}-X^++X^--Y^++Y^-\),\nn\\
J_2=\Delta+ig\(\frac{2}{\sqrt{X^-Y^-}}-\frac{2}{\sqrt{X^+Y^+}}+X^+-X^-+Y^+-Y^-\).
\eea
Then we also expand these two charges in $\de$ up to $\ml{O}(\de^3)$ as
\be
J_1=J_1^{(0)}+J_1^{(1)}\de+J_1^{(2)}\de^2+\ml{O}(\de^3),\qquad J_2=J_2^{(0)}+J_2^{(1)}\de+J_2^{(2)}\de^2+\ml{O}(\de^3).
\ee
The coefficients have already been compute in \cite{Abbott:2009um}, and we report the results here. The result for $J_1$ is
\bea
&&J_1^{(0)}=4g\frac{r^2-1}{r}\sin\(\frac{\ti{p_0}}{4}\),\nn\\
&&J_1^{(1)}=\frac{2g}{r}\[r^2\cos\(\frac{\ti{p_0}}{4}+\phi\)-\cos\(\frac{\ti{p_0}}{4}-\phi\)\],\nn\\
&&J_1^{(2)}=\frac{3g}{2r}\sin\(\frac{\ti{p_0}}{4}-2\phi\).
\eea
Similar result for $J_2$ reads
\bea
&&J_2^{(0)}=\Delta-4g\frac{r^2+1}{r}\sin\(\frac{\ti{p_0}}{4}\),\nn\\
&&J_2^{(1)}=-\frac{2g}{r}\[r^2\cos\(\frac{\ti{p_0}}{4}+\phi\)+\cos\(\frac{\ti{p_0}}{4}-\phi\)\],\nn\\
&&J_2^{(2)}=\frac{3g}{2r}\sin\(\frac{\ti{p_0}}{4}-2\phi\).
\eea
Then we find the corrections to $\ml{E}=\Delta-J_2$ begin at order $\de^2$
\be\label{EnergyCorrection}
\de\ml{E}=\Delta-J_2-\sqrt{J_1^2+64g^2\sin^2\frac{\ti{p}}{4}}
=-\frac{gr\cos(2\phi)}{r^2+1}\sin\(\frac{\ti{p}}{4}\)\de^2.
\ee
We now need to find out $\de$ by solving the constraint
\be
2\pi n_1=q_4(x+i0)-q_7(x-i0),\quad x\in \ml{C}_{47},
\ee
where $\ml{C}_{47}$ is the square-root branch cut connecting the 4th and 7th sheets as we are considering the $RP^3$ magnons, for which $[K_r,K_u,K_v]=[0,J_1,J_1]$. We are interested in the leading finite-size corrections, therefore we can evaluate at $x=X^+$ as
\be
2\pi n=\frac{2\al X^+}{(X^+)^2-1}+2G_{\text{finite}}(X^++i0)+2G_{\text{finite}}(X^+-i0)
+\pi\beta(J_1+ J_2)-p.
\ee
Solving this equation, we can fix $\de$ as
\be\label{de2}
\de^2=\pm 64\sin^2(\ti{p}/4)e^{i\pi(n_1-\beta J_2)-2i\phi}
\exp\(\frac{-i\Delta r/2g}{e^{i\ti{p}/4}r^2-e^{-i\ti{p}/4}}\).
\ee
To ensure the energy corrections being a real number, we must impose $\de$ to be real (for generic case the contributions at the $\ml{O}(\delta)$ order may be nonvanishing). Then the real part of $\de^2$  is
\bea
&&\Re(\de^2)=|\de^2|=64\sin^2(\ti{p}/4)\exp\[-\frac{\Delta r(r^2+1)\sin(\ti{p}/4)}
{2g((r^2-1)^2+4r^2\sin^2(\ti{p}/4))}\]\nn\\
&&=64\sin^2(\ti{p}/4)\exp\[-\frac{2\(J_2+\sqrt{J_1^2+64g^2\sin^2(\ti{p}/4)}\)
\sqrt{J_1^2+64g^2\sin^2(\ti{p}/4)}\sin^2(\ti{p}/4)}
{J_1^2+64g^2\sin^4(\ti{p}/4)}\].
\eea
Inserting this expression into eq.~(\ref{EnergyCorrection}), we find
\bea
\de\ml{E}&=&\frac{-256g^2\sin^4(\ti{p}/4)}{\sqrt{J_1^2+64g^2\sin^2(\ti{p}/4)}}\cos(2\phi)\nn\\
&\times&\exp\[-\frac{2\(J_2+\sqrt{J_1^2+64g^2\sin^2(\ti{p}/4)}\)
\sqrt{J_1^2+64g^2\sin^2(\ti{p}/4)}\sin^2(\ti{p}/4)}
{J_1^2+64g^2\sin^4(\ti{p}/4)}\].
\eea
The condition $\Im(\de)=0$ gives the relation of the phase $\phi$ and the conserved charges
\be
2\phi=\pi n_1^\prime-\(\pi\beta+J_1\frac{\sin(\ti{p}/2)}{J_1^2+64g^2\sin^4(\ti{p}/4)}\)J_2
-J_1\frac{\sin(\ti{p}/2)\sqrt{J_1^2+64g^2\sin^2(\ti{p}/4)}}{J_1^2+64g^2\sin^4(\ti{p}/4)},
\ee
where $n_1^\prime=n_1$ for plus sign in (\ref{de2}) while $n_1^\prime=n_1-1$ for minus sign in (\ref{de2}). Here we still have $\ti{p}=p-2\pi\beta J_1$ as before.
\section{Conclusion}
In this note, we have constructed the classical spectral curve of $\ga$-deformed $\text{AdS}_4/\text{CFT}_3$, which is consistent with the twisted ABA equations at strong coupling. The $\ga$-deformed algebraic curve is not very different from the original one as we only adding some appropriate phases which can be easily obtained from the twisted generating functional and some other natural requirement. To check our proposal, we compute the finite-size corrections of the dyonic giant magnons in $\text{AdS}_4\times\text{CP}^3_{\beta}$ using our twisted algebraic curve. Our result is the same as the one in \cite{Ahn:2011nm}. We hope this can be further verified by construct the twisted S-matrix and appropriate twisted boundary as in \cite{Ahn:2010ws} for $\text{AdS}_5/\text{CFT}_4$, and using the generalized L\"uscher formula \cite{Ahn:2012hsa} to compute the corrections of the dispersion relation.
\par Recently, the quantum spectral curve for $\ga$-deformed $\text{AdS}_5/\text{CFT}_4$ has been construct in \cite{Kazakov:2015efa}, it's also very interesting to find out it for $\text{AdS}_4/\text{CFT}_3$ with $\ga$-deformation.

\section*{Acknowledgments}
We would like to thank Hao Ouyang and Chao-Guang Huang for very helpful discussions. This work was in part supported by Natural Science Foundation of China under Grant Nos. 11575202(HC, JW), 11275207(HC), 11690022(HC).

\begin{appendix}
\section{Useful formula in scaling limit}\label{appena}
We give some useful formula of quantities expanding in the scaling limit.
\bea
&&\frac{Q^+_a}{Q^-_a}\simeq\frac{Q^{++}_a}{Q_a}\simeq\exp[i(H_a(x)+\bar{H}_a(x))],\qquad
\frac{Q^-_a}{Q^+_a}\simeq\frac{Q^{--}_a}{Q_a}\simeq\exp[-i(H_a(x)+\bar{H}_a(x))],\nn\\
&&\frac{R^{(+)-}}{R^{(-)-}}\simeq\frac{R^{(+)+}}{R^{(-)+}}\simeq\exp[i(G_4(x)+G_{\bar4}(x)],\qquad
\frac{B^{(+)+}}{B^{(-)+}}\simeq\frac{B^{(+)-}}{B^{(-)-}}\simeq\exp[i(\bar{G}_4(x)+\bar{G}_{\bar4}(x))],\nn\\
&&\frac{1}{i}\sum_{j=1}^{K_4}\log\sigma_{BES}(u,u_{4,j})\simeq\frac{1}{2g}\sum_{j=1}^{K_4}\frac{x-x_{4,j}}{(x^2-1)(xx_{4,j}-1)(x_{4,j}^2-1)}
=-\bar{H}_4(x)-\frac{G_4(0)}{x^2-1},\nn\\
&&G_4(x)+G_{\bar4}(x)=\frac{\ml{Q}_1+x\ml{Q}_2}{x^2-1}+H_4(x)+H_{\bar4}(x),\nn\\
&&\bar{G}_4(x)+\bar{G}_{\bar4}(x)=-\frac{x^2\ml{Q}_1+x\ml{Q}_2}{x^2-1}+\bar{H}_4(x)+\bar{H}_{\bar4}(x).
\eea
\section{Global charges}\label{appenb}
The Dynkin labels of a state can be read off from the Bethe equations as follows: we expand the right hand side of the Bethe equations for each flavor of root at infinity while keeping other roots finite. In this way, one can obtain the Dynkin label $r_j$ via comparing this expansion with  $1-ir_j/h(\lambda)x_{j,k}+\ml{O}(1/x^2_{j,k})$ \cite{Beisert:2005fw}.
\bea
&&r_1=-\eta K_2-\eta\de D,\nn\\
&&r_2=2\eta K_2-\eta K_3-\eta K_1,\nn\\
&&r_3=-\eta K_2+\eta K_4+\eta K_{\bar4}+\eta\de D,\nn\\
&&r_4=L-(1+\eta)K_4+(1-\eta)K_{\bar4}+\eta K_3+\frac12(1-\eta)\de D,\nn\\
&&r_{\bar4}=L-(1+\eta)K_{\bar4}+(1-\eta)K_4+\eta K_3+\frac12(1-\eta)\de D.
\eea
For $su_2$ grading discussed in the main text and the $SU(4)$ Dynkin label $[p_1,q,p_2]$, we have
\bea
&&q=r_1+r_2+r_3=K_4+K_{\bar4}-K_3-K_1,\nn\\
&&p_1=r_4=L-2K_4+K_3,\nn\\
&&p_2=r_{\bar4}=L-2K_{\bar4}+K_3.
\eea
\end{appendix}


\begin{thebibliography}{}
\bibitem{Lukowski:2008eq}
  T.~Lukowski and O.~Ohlsson Sax,
  ``Finite size giant magnons in the $SU(2) \times SU(2)$ sector of $AdS_4 \times CP^3$,''
  JHEP {\bf 0812} (2008) 073
  doi:10.1088/1126-6708/2008/12/073
  [arXiv:0810.1246 [hep-th]].
\bibitem{Ahn:2011nm}
  C.~Ahn and P.~Bozhilov,
  ``Finite-size Giant Magnons on $AdS_4 \times CP^3_{\gamma}$,''
  Phys.\ Lett.\ B {\bf 703} (2011) 186
  doi:10.1016/j.physletb.2011.07.065
  [arXiv:1106.3686 [hep-th]].
\bibitem{Maldacena:1997re}
  J.~M.~Maldacena,
  ``The Large N limit of superconformal field theories and supergravity,''
  Int.\ J.\ Theor.\ Phys.\  {\bf 38} (1999) 1113
   [Adv.\ Theor.\ Math.\ Phys.\  {\bf 2} (1998) 231]
  doi:10.1023/A:1026654312961
  [hep-th/9711200].
\bibitem{Minahan:2002ve}
  J.~A.~Minahan and K.~Zarembo,
  ``The Bethe ansatz for $\ml{N}=4$ superYang-Mills,''
  JHEP {\bf 0303} (2003) 013
  doi:10.1088/1126-6708/2003/03/013
  [hep-th/0212208].
\bibitem{Bena:2003wd}
  I.~Bena, J.~Polchinski and R.~Roiban,
  ``Hidden symmetries of the $AdS_5\times S^5$ superstring,''
  Phys.\ Rev.\ D {\bf 69} (2004) 046002
  doi:10.1103/PhysRevD.69.046002
  [hep-th/0305116].
\bibitem{Minahan:2008hf}
  J.~A.~Minahan and K.~Zarembo,
  ``The Bethe ansatz for superconformal Chern-Simons,''
  JHEP {\bf 0809} (2008) 040
  doi:10.1088/1126-6708/2008/09/040
  [arXiv:0806.3951 [hep-th]].
\bibitem{Arutyunov:2008if}
  G.~Arutyunov and S.~Frolov,
  ``Superstrings on $AdS_4\times CP^3$ as a Coset Sigma-model,''
  JHEP {\bf 0809} (2008) 129
  doi:10.1088/1126-6708/2008/09/129
  [arXiv:0806.4940 [hep-th]].
\bibitem{Beisert:2010jr}
  N.~Beisert {\it et al.},
  ``Review of AdS/CFT Integrability: An Overview,''
  Lett.\ Math.\ Phys.\  {\bf 99} (2012) 3
  doi:10.1007/s11005-011-0529-2
  [arXiv:1012.3982 [hep-th]].
\bibitem{Leigh:1995ep}
  R.~G.~Leigh and M.~J.~Strassler,
  ``Exactly marginal operators and duality in four-dimensional N=1 supersymmetric gauge theory,''
  Nucl.\ Phys.\ B {\bf 447} (1995) 95
  doi:10.1016/0550-3213(95)00261-P
  [hep-th/9503121].
\bibitem{Lunin:2005jy}
  O.~Lunin and J.~M.~Maldacena,
  ``Deforming field theories with $U(1)\times U(1)$ global symmetry and their gravity duals,''
  JHEP {\bf 0505} (2005) 033
  doi:10.1088/1126-6708/2005/05/033
  [hep-th/0502086].

\bibitem{Berenstein:2004ys}
  D.~Berenstein and S.~A.~Cherkis,
  ``Deformations of $\ml{N}=4$ SYM and integrable spin chain models,''
  Nucl.\ Phys.\ B {\bf 702}, 49 (2004)
   doi:10.1016/j.nuclphysb.2004.09.005
  [hep-th/0405215].

\bibitem{Beisert:2005if}
  N.~Beisert and R.~Roiban,
  ``Beauty and the twist: The Bethe ansatz for twisted $\ml{N}=4$ SYM,''
  JHEP {\bf 0508}, 039 (2005)
  doi:10.1088/1126-6708/2005/08/039
  [hep-th/0505187].  

\bibitem{Imeroni:2008cr}
  E.~Imeroni,
  ``On deformed gauge theories and their string/M-theory duals,''
  JHEP {\bf 0810}, 026 (2008)
  doi:10.1088/1126-6708/2008/10/026
  [arXiv:0808.1271 [hep-th]]

\bibitem{Giataganas:2013dha}
  D.~Giataganas, L.~A.~Pando Zayas and K.~Zoubos,
  ``On Marginal Deformations and Non-Integrability,''
  JHEP {\bf 1401}, 129 (2014)
  doi:10.1007/JHEP01(2014)129
  [arXiv:1311.3241 [hep-th]].

\bibitem{Chen:2016geo}
  H.~H.~Chen, P.~Liu and J.~B.~Wu,
  ``Y-system for $\gamma$-deformed ABJM Theory,''
  [arXiv:1611.02804 [hep-th]].


\bibitem{Shenderovich:2008bs}
  I.~Shenderovich,
  ``Giant magnons in AdS(4) / CFT(3): Dispersion, quantization and finite-size corrections,''
  arXiv:0807.2861 [hep-th].



\bibitem{Frolov:2005dj}
  S.~Frolov,
  ``Lax pair for strings in Lunin-Maldacena background,''
  JHEP {\bf 0505} (2005) 069
  doi:10.1088/1126-6708/2005/05/069
  [hep-th/0503201].
\bibitem{Schimpf:2009rk}
  M.~Schimpf and R.~C.~Rashkov,
  ``A Note on strings in deformed $AdS_4\times CP^3$: Giant magnon and single spike solutions,''
  Mod.\ Phys.\ Lett.\ A {\bf 24} (2009) 3227
  doi:10.1142/S0217732309032113
  [arXiv:0908.2246 [hep-th]].
\bibitem{Gromov:2009at}
  N.~Gromov and F.~Levkovich-Maslyuk,
  ``Y-system, TBA and Quasi-Classical strings in AdS(4) x CP3,''
  JHEP {\bf 1006} (2010) 088
  doi:10.1007/JHEP06(2010)088
  [arXiv:0912.4911 [hep-th]].
\bibitem{Kim:2014vka}
  M.~Kim,
  ``Spectral curve for $\gamma$-deformed $AdS/CFT$,''
  Phys.\ Lett.\ B {\bf 735} (2014) 332
  doi:10.1016/j.physletb.2014.06.052
  [arXiv:1401.4032 [hep-th]].
\bibitem{Gromov:2008ie}
  N.~Gromov, S.~Schafer-Nameki and P.~Vieira,
  ``Quantum Wrapped Giant Magnon,''
  Phys.\ Rev.\ D {\bf 78} (2008) 026006
  doi:10.1103/PhysRevD.78.026006
  [arXiv:0801.3671 [hep-th]].

\bibitem{Astolfi:2007uz}
  D.~Astolfi, V.~Forini, G.~Grignani and G.~W.~Semenoff,
  ``Gauge invariant finite size spectrum of the giant magnon,''
  Phys.\ Lett.\ B {\bf 651}, 329 (2007)
  doi:10.1016/j.physletb.2007.06.002
  [hep-th/0702043].

\bibitem{Ramadanovic:2008qd}
  B.~Ramadanovic and G.~W.~Semenoff,
  ``Finite Size Giant Magnon,''
  Phys.\ Rev.\ D {\bf 79}, 126006 (2009)
  doi:10.1103/PhysRevD.79.126006
  [arXiv:0803.4028 [hep-th]].


\bibitem{Grignani:2008te}
  G.~Grignani, T.~Harmark, M.~Orselli and G.~W.~Semenoff,
  ``Finite size Giant Magnons in the string dual of N=6 superconformal Chern-Simons theory,''
  JHEP {\bf 0812}, 008 (2008)
  doi:10.1088/1126-6708/2008/12/008
  [arXiv:0807.0205 [hep-th]].

\bibitem{c1} M.~Benna, I.~Klebanov, T.~Klose and M.~Smedback,
  ``Superconformal Chern-Simons Theories and AdS(4)/CFT(3) Correspondence,''
  JHEP {\bf 0809}, 072 (2008)
  doi:10.1088/1126-6708/2008/09/072
  [arXiv:0806.1519 [hep-th]].

\bibitem{Terashima:2008ba}
  S.~Terashima and F.~Yagi,
  ``Orbifolding the Membrane Action,''
  JHEP {\bf 0812}, 041 (2008)
  doi:10.1088/1126-6708/2008/12/041
  [arXiv:0807.0368 [hep-th]].

  \bibitem{Imamura:2008ji}
  Y.~Imamura and S.~Yokoyama,
  ``N=4 Chern-Simons theories and wrapped M-branes in their gravity duals,''
  Prog.\ Theor.\ Phys.\  {\bf 121}, 915 (2009)
  doi:10.1143/PTP.121.915
  [arXiv:0812.1331 [hep-th]].

  \bibitem{Berenstein}
  D.~Berenstein and M.~Romo,
  ``Aspects of ABJM orbifolds,''
  Adv.\ Theor.\ Math.\ Phys.\  {\bf 14}, no. 6, 1717 (2010)
  [arXiv:0909.2856 [hep-th]].

\bibitem{Bai:2016pxs}
  N.~Bai, H.~H.~Chen, X.~C.~Ding, D.~S.~Li and J.~B.~Wu,
  ``Integrability of Orbifold ABJM Theories,''
  JHEP {\bf 1611} (2016) 101
  doi:10.1007/JHEP11(2016)101
  [arXiv:1607.06643 [hep-th]].
\bibitem{Abbott:2010yb}
  M.~C.~Abbott, I.~Aniceto and D.~Bombardelli,
  ``Quantum Strings and the $AdS_4/CFT_3$ Interpolating Function,''
  JHEP {\bf 1012} (2010) 040
  doi:10.1007/JHEP12(2010)040
  [arXiv:1006.2174 [hep-th]].

\bibitem{Abbott:2013mpa}
  M.~C.~Abbott,
  ``The $AdS_{3}\times S^{3} \times S^{3}\times S^{1}$ Hernandez-Lopez phases: a semiclassical derivation,''
  J.\ Phys.\ A {\bf 46}, 445401 (2013)
  doi:10.1088/1751-8113/46/44/445401
  [arXiv:1306.5106 [hep-th]].
\bibitem{Minahan:2006bd}
  J.~A.~Minahan, A.~Tirziu and A.~A.~Tseytlin,
  ``Infinite spin limit of semiclassical string states,''
  JHEP {\bf 0608} (2006) 049
  doi:10.1088/1126-6708/2006/08/049
  [hep-th/0606145].
\bibitem{Sax:2008in}
  O.~Ohlsson Sax,
  ``Finite size giant magnons and interactions,''
  Acta Phys.\ Polon.\ B {\bf 39} (2008) 3143
  [arXiv:0810.5236 [hep-th]].

\bibitem{Ahn:2010ws}
  C.~Ahn, Z.~Bajnok, D.~Bombardelli and R.~I.~Nepomechie,
  ``Twisted Bethe equations from a twisted S-matrix,''
  JHEP {\bf 1102} (2011) 027
  doi:10.1007/JHEP02(2011)027
  [arXiv:1010.3229 [hep-th]].
\bibitem{Abbott:2009um}
  M.~C.~Abbott, I.~Aniceto and O.~Ohlsson Sax,
  ``Dyonic Giant Magnons in $CP^3$: Strings and Curves at Finite $J$,''
  Phys.\ Rev.\ D {\bf 80} (2009) 026005
  doi:10.1103/PhysRevD.80.026005
  [arXiv:0903.3365 [hep-th]].

\bibitem{Ahn:2012hsa}
  C.~Ahn, D.~Bombardelli and M.~Kim,
  ``Finite-size effects of $\beta$-deformed $AdS_5/CFT_4$ at strong coupling,''
  Phys.\ Lett.\ B {\bf 710} (2012) 467
  doi:10.1016/j.physletb.2012.03.001
  [arXiv:1201.2635 [hep-th]].
\bibitem{Kazakov:2015efa}
  V.~Kazakov, S.~Leurent and D.~Volin,
  ``T-system on T-hook: Grassmannian Solution and Twisted Quantum Spectral Curve,''
  arXiv:1510.02100 [hep-th].

\bibitem{Beisert:2005fw}
  N.~Beisert and M.~Staudacher,
  ``Long-range $psu(2,2|4)$ Bethe Ansatze for gauge theory and strings,''
  Nucl.\ Phys.\ B {\bf 727} (2005) 1
  doi:10.1016/j.nuclphysb.2005.06.038
  [hep-th/0504190].
\end{thebibliography}
\end{document}